\documentstyle[osa,manuscript,graphicx]{revtex}

\begin{document}

\def\lsim{\ ^<\llap{$_\sim$}\ }
\def\gsim{\ ^>\llap{$_\sim$}\ }
\def\r2{\sqrt 2}
\def\beq{\begin{equation}}
\def\eeq{\end{equation}}
\def\beqn{\begin{eqnarray}}
\def\eeqn{\end{eqnarray}}
\def\rmuu{\gamma^{\mu}}
\def\rmud{\gamma_{\mu}}
\def\PL{{1-\gamma_5\over 2}}
\def\PR{{1+\gamma_5\over 2}}
\def\sinW2{\sin^2\theta_W}
\def\AEM{\alpha_{EM}}
\def\mul{M_{\tilde{u} L}^2}
\def\mur{M_{\tilde{u} R}^2}
\def\mdl{M_{\tilde{d} L}^2}
\def\mdr{M_{\tilde{d} R}^2}
\def\mz2{M_{z}^2}
\def\c2b{\cos 2\beta}
\def\au{A_u}
\def\ad{A_d}
\def\cob{\cot \beta}
\def\v#1{v_#1}
\def\tb{\tan\beta}
\def\epem{$e^+e^-$}
\def\KK{$K^0$-$\bar{K^0}$}
\def\wi{\omega_i}
\def\xj{\chi_j}
\def\Wmu{W_\mu}
\def\Wnu{W_\nu}
\def\m#1{{\tilde m}_#1}
\def\mH{m_H}
\def\mw#1{{\tilde m}_{\omega #1}}
\def\mx#1{{\tilde m}_{\chi^{0}_#1}}
\def\mc#1{{\tilde m}_{\chi^{+}_#1}}
\def\mwi{{\tilde m}_{\omega i}}
\def\mxi{{\tilde m}_{\chi^{0}_i}}
\def\mci{{\tilde m}_{\chi^{+}_i}}
\def\mz{M_z}
\def\sw{\sin\theta_W}
\def\cw{\cos\theta_W}
\def\cb{\cos\beta}
\def\sb{\sin\beta}
\def\rwi{r_{\omega i}}
\def\rxj{r_{\chi j}}
\def\rfp{r_f'}
\def\Kik{K_{ik}}
\def\Fq2{F_{2}(q^2)}
%

\begin{center}{\Large \bf  
Out-going Muon Flux from Neutralino Annihilation in the Sun and the  Earth
in Supergravity Unification } \\
\vskip.25in
{Achille Corsetti and  Pran Nath  }

{\it
 Department of Physics, Northeastern University, Boston, MA 02115-5005, USA\\
}

\end{center}

\begin{abstract}  
Predictions for the out-going  muon fluxes  
from the annihilation of neutralinos in the center of 
the Sun and the Earth  in supergravity unification are given.
Effects of uncertainties of the input data are analysed. 
 It is shown that the out-going muon flux measurements from
the Sun and the Earth are complementary, with the Earth providing 
a larger flux for low  values of fine tuning and the Sun providing 
a larger flux for high values of fine tuning,   
and that only a combination of the two  can compete favorably 
with the direct detection measurements. 
The predictions are compared with the recent limits
from BAIKAL NT-96, MACRO, and BAKSAN.  
\end{abstract}

Recently analyses of the data from several neutrino 
telescopes
(BAIKAL NT-96\cite{nt96},MACRO\cite{macro},
BAKSAN\cite{baksan}) 
 have produced 
bench mark limits on the out-going muon fluxes from the annihilation
of dark matter wimps in the centers of the Sun and the  
Earth \cite{silk,kamion,jungman}. 
In this Letter we give predictions for these out-going muon fluxes 
in supergravity grand unification\cite{chams,applied}
taking into account the uncertainties 
in the local wimp density as well as the uncertainties in the rms wimp velocity
and we compare these predictions with the recent experimental  
 limits\cite{nt96,macro,baksan}. 
We study the relative merit of using the Sun vs the Earth
for the detection of dark matter as the SUSY spectrum gets progressively
 heavier and  the fine tuning parameter of the 
radiative electro-weak symmetry
breaking gets  progressively  larger. We also make a similar comparison 
of the direct vs the indirect detection of neutralino dark matter and show 
that only a combination of the out-going flux measurements from the
Sun and the Earth can compete effectively with the direct detection of 
neutralino dark matter in mapping the parameter space of minimal 
supergravity.

Supergravity unified models are extensions of the Standard Model, 
consistent with all current data, and under the constraint of R partiy 
invariance lead to the existence of a cosmologically stable neutralino
$\tilde\chi_1^0$ which can act as a candidate for cold dark matter. 
  In these models the breaking of supersymmetry
  takes place in the hidden sector and the breaking is communicated via 
gravitational interactions to the visible sector where 
the breaking of supersymmetry can be parametrized by the 
following set for the minimal case (mSUGRA)\cite{applied}: 
the universal scalar mass ($m_0$) at 
the GUT scale, the universal gaugino mass ($m_{1/2}$) at the GUT scale, 
the universal trilinear  coupling $A_0$ at the GUT scale, 
$tan\beta=<H_2>/<H_1>$, where $<H_2>$ gives mass to the 
up quarks and $<H_1>$ gives mass to the down quarks and the leptons,
and sign($\mu$) where $\mu H_1H_2$ is the Higgs mixing term in the 
superpotential. Detailed relic density analyses show that supergravity 
models can produce a wimp relic density consistent with a range
 of current cosmological models.  We shall assume this relic density
 to lie in the range $0.05\leq \Omega_{\tilde\chi_1^0} h^2\leq 0.3$, where 
 $\Omega_{\tilde\chi_1^0}=\rho_{\tilde\chi_1^0}/\rho_c$, 
 with $\rho_{\tilde\chi_1^0}$ 
  the average density of neutralinos in the universe and
  $\rho_c=3H^2/8\pi G_N$  the critical matter density, where $G_N$ 
  is the Newtonian constant and h is the Hubble parameter in units of
  $100kmsec^{-1}Mpc^{-1}$. 

The basic elements of the analysis of the physics of out-going muons 
are the following: 
The Milky Way neutralinos are trapped by scattering 
and gravitational pull by the Earth and the Sun and accumulate in their 
cores where they eventually annihilate. Some of the final annihilation remnants 
contain neutrinos which propagate and undergo charge current interactions
in the rock surrounding the detector and produce out-going 
muons which are detected. The basic theory of the various 
elements of the analysis are fairly well laid out 
and we follow the standard literature here\cite{silk,kamion}.  
However,  we point out some basic features  where specific 
elements of the SUGRA models enter. The flux of the out-going 
muons is given by $\Phi_{\mu}=\Gamma_A f$, where $\Gamma_A$ is 
the rate of $\tilde\chi_1^0-\tilde\chi_1^0$ annihilation in the center 
of the Earth or the Sun and f is the product of remaining 
factors\cite{silk,kamion,jungman}.
A significant portion of the SUSY dependence in contained in 
$\Gamma_A$ which is given analytically by\cite{gs} 
\begin{equation}
\Gamma_A=\frac{C}{2}tanh^2(t/\tau) 
\end{equation}
Here C is the capture rate of the neutralinos which depends on the
wimp relic density and on the wimp-nucleus cross section, t is
the age of the Earth or the Sun and $\tau=(CC_A)^{-1/2}$ where
$C_A$ depends on the wimp annihilation cross-section. 
In the limit $t>>\tau$, one finds $\Gamma_A\sim C/2$, i.e., 
an equilibrium is reached between annihilation and capture.
The condition of equilibrium depends on the particulars of the
SUSY parameter space. For the case of the Sun one finds that
the equilibrium condition is realized over essentially 
all of the parameter space of the model while for the case 
of the Earth this condition is highly model dependent. 

The predictions of the out-going muon fluxes in supergravity unifed 
models will be constrained by several factors. In addition to the radiative
electro-weak symmetry breaking\cite{applied} and the relic density
 constraints we impose the constraint from the flavor changing
neutral current process $b\rightarrow s+\gamma$. 
This process is known to constrain dark matter analyses severely
by eliminating large segments of the mSUGRA parameter 
space via the relic density constraint\cite{bsgamma}.  The parameter
space of mSUGRA  will be further constrained by the naturalness 
criterion which we take to imply $m_0\leq 1$ TeV, $m_{\tilde g}\leq 1$ TeV,
$tan\beta \leq 25$, and $-7\leq (A_t/m_0)\leq 7$,
where $A_t$ is the value of $A_0$ at the electro-weak scale for the top
quark channel.
In the analysis of the relic density we use the accurate
method which includes the correct thermal averaging over the Breit-Wigner
poles in the neutralino s channel annihilation\cite{accurate}.

In the analysis below we study the dependence of our predictions for 
the out-going muon fluxes from 
the Earth and the Sun  
 including the uncertainties of the neutralino relic density in the Milky
Way. Current estimates show that the local wimp density
can vary in the range\cite{gates}  $(0.2-0.7) GeVcm^{-3}$.
We parametrize this variation by $\xi=\rho_{\tilde\chi_1^0}/\rho_0$
which for  $\rho_0=0.3 GeVcm^{-3}$ lies in the range 
$0.7\leq \xi\leq 2.3$.  Further, we study the dependence of 
the out-going muon fluxes on the uncertainties in the wimp velocity\cite{knapp}
 the details of which are given below.
There could also be effects on the out-going muon fluxes due to
rotation of the galaxy\cite{kk}.  However, such effects may be 
small (e.g. an analysis of these effect on the direct
detection rates of dark matter was found to be only 
$\sim 10\%$). A brief discussion of the effects on the prediction of the 
muon fluxes under the constraint of the annual modulation
signal claimed by the DAMA Collaboration\cite{dama} is also given.
Finally, we give an analysis of the out-going muon flux from the Sun
and the Earth as a function of the fine tuning parameter and discuss
the relative merits of the direct vs the indirect detection. 
In the analysis below  we neglect the $\nu$ oscillation effects. These effects 
 could modify the detection rates from $10\%-50\%$ but are highly model
 dependent on the nature of the assumed $\nu$ oscillation\cite{fornengo}.
 Inclusion of these effects would not change the conclusions of this work.
 
We discuss now the results of our analysis. 
In Fig.(1a) a plot of the maximum out-going muon flux 
 for the case of
the Earth is given as a function of $m_{\tilde\chi_1^0}$
 for three $\xi$
values: $\xi=0.7$, $\xi=1$ and $\xi=2.3$.
One finds that 
the maximum out-going muon flux is generally peaked for $m_{\tilde\chi_1^0}$
below 
 60 GeV after which it shows a fall off with increasing
  $m_{\tilde\chi_1^0}$.  
Interestingly one finds that the MACRO
limits are already stringent enough to put constraints on mSUGRA  
in this case although the part of the parameter space 
so constrained is  rather small. 
In Fig.(1b) a similar plot  is given for the case of the out-going muon
flux from the Sun.
  However, here the MACRO limits 
do not put any constraint on mSUGRA. In fact,
the experimental sensitivity will have to increase by a factor
of 10 or more before mSUGRA can be constrained
 by the Sun flux. A comparison of Fig.(1a) and Fig.(1b) shows that for 
  relatively low values of $m_{\tilde\chi_1^0}$ the Earth provides a better
source for wimp detection relative to the 
Sun. We shall return to a more detailed comparison of the goodness of the 
 Earth vs the Sun for wimp detection  in the  discussion of   
Fig.3. 
  
  The analysis of Figs.(1a) and Fig.(1b) was based on a Maxwellian velocity 
distribution of the wimps with an rms wimp velocity of 
$v=270 ~km/s$. However, there is the possibility of a significant
uncertainty in the rms wimp velocity 
with estimates of the uncertainty ranging from $\pm 24$ km/s to 
$\pm 70$ km/s \cite{knapp}.
 For illustrative purposes we study the effects 
for the range $v=270\pm 50 ~km/s$. The results are shown in Fig.(2a) 
for the maximum out-going muon flux for the Earth and in Fig.(2b) for
the maximum out-going muon flux for the Sun. Here
we find that the $\pm 50$ km/s variation of the rms wimp velocity can
generate significant variations in the out-going muon fluxes. 
 Next we discuss the effect on our analysis of including the constraints 
 that will allow the annual modulation signal in the direct detection
 of wimps claimed by the DAMA Collaboration\cite{dama}.
The constraint that the supergravity unified models 
generate $\tilde\chi_1^0$-proton cross-sections compatible with 
the annual modulation signal observed by DAMA has been analysed
elsewhere\cite{bottino,milky1} and the reduced set of mSUGRA configurations 
compatible 
with the claimed DAMA signal have been identified\cite{bottino,milky1}.  
An analysis including the DAMA constraint
 shows that the peak values of the maximum out-going muon fluxes 
for the Earth and the Sun are not significantly affected.
Thus at least in the peak region the MACRO limits for the out-going 
muon flux for the Earth  put a further
constraint on the reduced set of configurations allowed by the 
DAMA cut\cite{bottino,milky1}.

In Fig.(3a) we study the mean out-going  muon flux gotten by averaging
 over the allowed parameter
space for fixed fine tuning $\Phi$  as a function of 
$\sqrt\Phi$. Roughly speaking the fine tuning 
parameter measures how heavy the SUSY spectrum is and we use here
the criterion given in Ref.\cite{lung} which defines $\Phi$=$\frac{1}{4}$
+$\frac{\mu^2}{M_Z^2}$. 
Fig.(3a) shows that the mean out-going muon  
flux decreases sharply with $\sqrt\Phi$. Remarkably one finds that  
   the Earth flux falls by many more decades than the Sun flux
   over the allowed range of $\sqrt\Phi$. Thus for 
   low values of $\sqrt\Phi$, i.e., $\sqrt\Phi \leq 2.5$
    (which imply a relatively light SUSY spectrum) 
   the Earth flux would 
    be more easily accessible to neutrino telescope searches, while 
   an opposite situation holds for larger $\sqrt\Phi$ values,
   i.e., $\sqrt\Phi \geq 2.5$ (which imply a
   relatively heavy SUSY spectrum).
   This means that for a low SUSY mass spectrum,
   e.g., with values near the current experimental lower limits 
   from accelerators, the prime source for the discovery of dark
   matter wimps for neutrino telescopes is the out-going muon
   flux from the Earth.
    However, as the lower
   limits on the SUSY masses at accelerators increase and approach
   their naturalness limits, the out-going muon flux from the Sun
   will become the prime
   source for the discovery of dark matter wimps for neutrino telescopes. 
   Of course, as
   the SUSY spectrum becomes progressively heavier, one would need
   progressively larger sensitivities in neutrino telescopes to probe
   the parameter space remaining in supergravity models. 
 
  	Finally we discuss the relative merits of the indirect vs
  	the direct
  	detection (For the details of computation of direct detection rates 
  	see, e.g., Ref.\cite{direct}).
  	In Fig.(3b) the maximum and the minimum 
  	out-going muon flux from the Sun 
  	and the Earth is plotted against the direct detection  rate
  	for Ge.
  	The analysis shows that while the direct detection rate spans
  	 a wide range, i.e., roughly about 4 decades, this range is  
  	 much smaller than the range of the out-going muon flux  
  	 from the Earth. However, the range of the outgoing muon flux 
  	 from the Sun is comparable to the range of the direct detection 
         rate. The analysis implies that the 
 	direct detection will certainly be superior 
 	to measurements from neutrino telescopes using only the 
 	out-going muon flux from the Earth 
 	at least for relatively larger fine 
 	tunings (i.e., $\Phi >2.5$, see Fig.(3a)).
	However,  for relatively larger fine tunings the out-going muon
	flux
  	measurements from the Sun will compete favorably with the direct
  	detection measurements. 

  	In summary our conclusion is that 
  	neutrino telescopes measuring the out-going muon
  	flux only for the Earth or only for the Sun cannot 
  	compete with the direct dark matter searches in 
  	mapping the full parameter space of mSUGRA  within the
  	assumed naturalness limits. However, 
  	the out-going muon flux measurements from both the Earth and 
  	the Sun can  
  	be competitive with the direct dark matter searches 
  	provided the improvement in the sensitivities of 
  	the indirect and the direct detection measurements keep pace, i.e.,
  	the sensitivity of measurements improves by the same number
  	of decades in each experiment.
  	We have also investigated the effects of non-universalities 
  	of the soft SUSY breaking parameters at the GUT scale on the
  	analysis presented here.
  	Our conclusions here are very similar to the one's
  	given above. The predictions of the muon fluxes given here
  	and their dependence on fine tuning will have implications for
  	other neutrino telescopes regarding their ability to search for
  	supersymmetric dark matter, e.g.,  AMANDA\cite{halzen}
  	 at the South Pole which is  collecting data and possibly other 
  	 future neutrino telescopes which are in the developmental stage.
  	
  This research was supported in part by NSF grant 
PHY-96020274.
 
\vspace{-12cm} 
\begin{figure}
\begin{center}
\includegraphics[angle=0,width=5.5in]{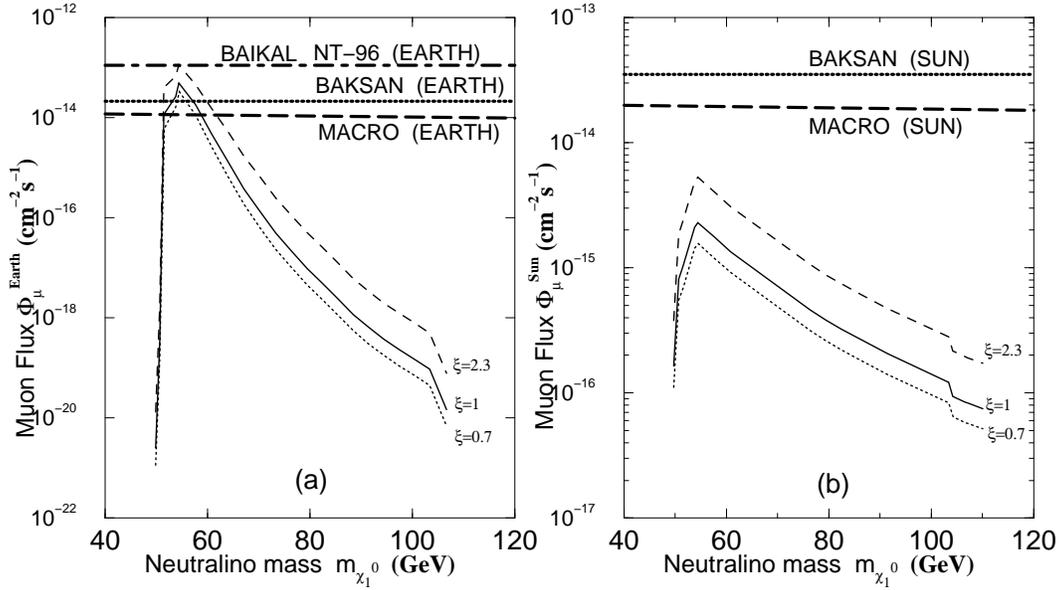}
\caption{(a): Plot of the maximum out-going muon flux 
 for the Earth in mSUGRA 
as a function of
the neutralino mass for three different values of the local wimp 
density corresponding to $\xi=0.7, 1, 2.3$, and $\mu>0$ (Our $\mu$
sign convention is as in Ref.[19]). The horizontal lines
show limits from experiment; (b) Same as (a) except
for the Sun.}
\end{center}
\end{figure}
\newpage  	
\begin{figure}
\begin{center}
\includegraphics[angle=0,width=5.5in]{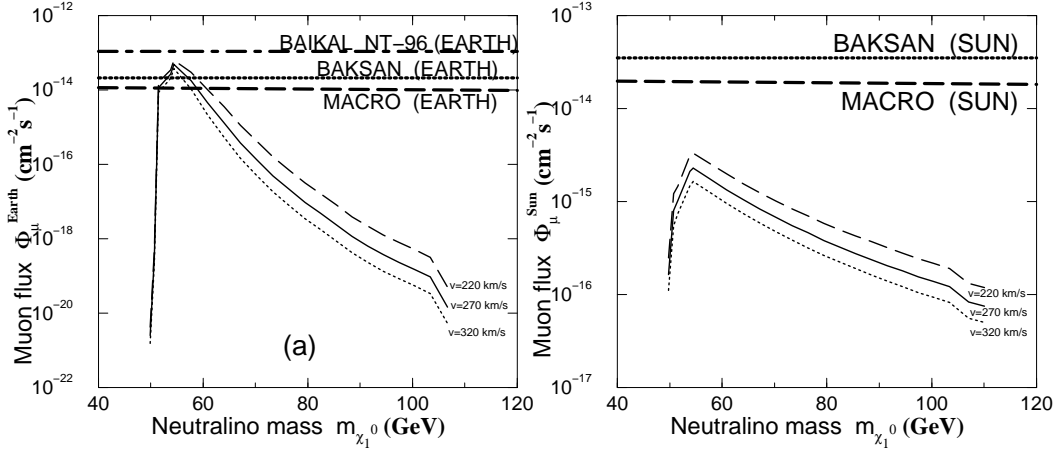}
\caption{ Same as Fig.(1a) except that the analysis is for
$\xi=1$ and for three different rms wimp velocities in the halo, i.e.,
v=220 km/s, 270 km/s, and 320 km/s.
(b) Same as (a) except for the Sun.}
\end{center}
\end{figure}

 
\begin{figure}
\begin{center}
\includegraphics[angle=0,width=5.5in]{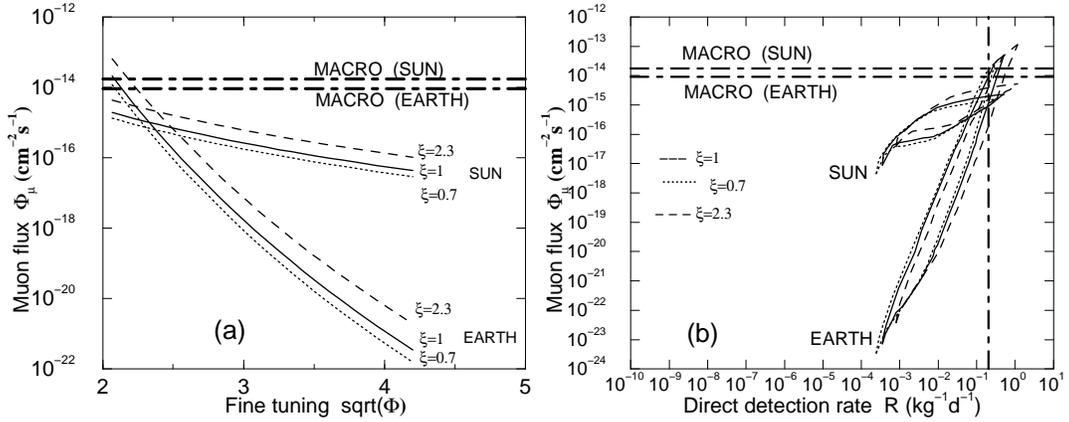}
\caption{(a): Plot of the mean out-going muon flux for the Sun and the Earth
 as a function of the fine tuning parameter $\sqrt\Phi$ for three values of 
 $\xi$. The experimental limits from MACRO, which are currently the best 
 limits, are shown by the horizontal lines; (b) Plot of the maximum and 
 the minimum of the out-going muon flux for the Sun and for the Earth as 
 a function of the direct detection rate for Ge. The verticle line is the 
 upper limit from Ref.[20].}
\end{center}
\end{figure}

\end{document}